\documentclass[aip,apl,reprint]{revtex4-1}

\usepackage{graphicx}
\usepackage{dcolumn}
\usepackage{bm}

\begin{document}

\title{Magnetic order near 270 K in mineral and synthetic Mn$_2$FeSbO$_6$ ilmenite}

\author{R. Mathieu}\email{roland.mathieu@angstrom.uu.se}
\affiliation{Department of Engineering Sciences, Uppsala University, Box 534, SE-751 21 Uppsala, Sweden}

\author{S. A. Ivanov}
\affiliation{Department of Engineering Sciences, Uppsala University, Box 534, SE-751 21 Uppsala, Sweden}
\affiliation{Department of Inorganic Materials, Karpov' Institute of Physical Chemistry, Vorontsovo pole, 10 105064, Moscow K-64, Russia}

\author{G. V. Bazuev}
\affiliation{Institute of Solid-State Chemistry, Ural Branch of the Russian Academy of Science, 620999 Ekaterinburg, GSP-145, Russia}

\author{M. Hudl}
\affiliation{Department of Engineering Sciences, Uppsala University, Box 534, SE-751 21 Uppsala, Sweden}

\author{P. Lazor}
\affiliation{Department of Earth Sciences, Uppsala University, Villav\"agen 16, SE-752 36 Uppsala, Sweden}

\author{I. V. Solovyev}	
\affiliation{National Institute for Materials Science, 1-2-1 Sengen, Tsukuba, Ibaraki 305-0047, Japan}

\author{P. Nordblad}
\affiliation{Department of Engineering Sciences, Uppsala University, Box 534, SE-751 21 Uppsala, Sweden}

\date{\today}

\begin{abstract}

The structural and magnetic properties of Mn$_2$FeSbO$_6$ single-crystalline mineral and ceramic samples synthesized under thermobaric treatment have been investigated, and compared to theoretical predictions based on first-principles electronic structure calculations. This ilmenite system displays a sharp magnetic transition just below the room temperature related to a ferrimagnetic ordering of the Mn$^{2+}$ and Fe$^{3+}$ cations, which makes Mn$_2$FeSbO$_6$ a promising candidate for designing functional magnetic materials.

\end{abstract}

\maketitle

There is an intense research focusing in finding magnetic functional materials to build more efficient and less costly appliances. Such materials include e.g. magnetoelectric\cite{me} and magnetocaloric\cite{mce} materials, to be used in spintronic application or magnetic refrigeration. Our planet contains many minerals comprising magnetic elements such as iron, which can be used to build such devices. Minerals of more complex oxides, synthesized in extreme conditions of temperature, pressure and time (billion years), can also be found. For example solid solutions of hematite-ilmenite Fe$_2$O$_3$-FeTiO$_3$ which naturally exist in the Earth's crust, were found to display exchange bias effect.\cite{fabian} Another interesting example is the mineral melanostibite Mn$_2$FeSbO$_6$, crystallizing with the ilmenite structure.\cite{melano} This mineral is mainly abundant in Sweden, and has largely unknown magnetic properties. 

In this letter, we report the magnetic properties of this natural material, as well as those of synthetic Mn$_2$FeSbO$_6$ samples. The small single-crystals of the mineral was found to display a sharp magnetic transition just below room-temperature (near $\sim$ 270K). By gradually improving the synthesis conditions, we have obtained nearly phase pure synthetic Mn$_2$FeSbO$_6$, with magnetic properties similar to those of the natural mineral. Our results suggest that functional magnetic materials such as magnetoelectric or magnetocaloric materials could be obtained using melanostibite as a basis.\\

\begin{figure}
\includegraphics{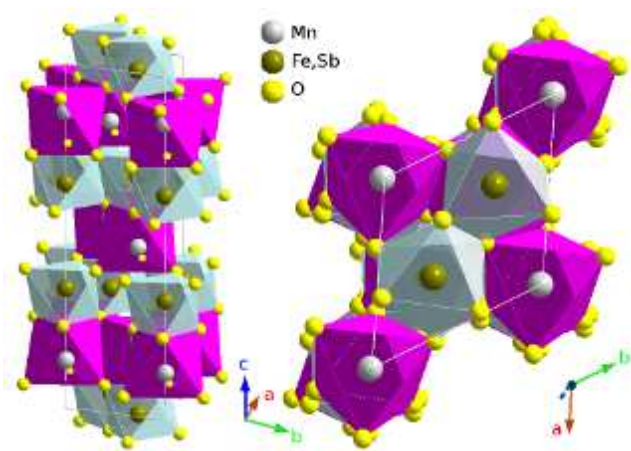}
\caption{(Color online) Different views of the octahedral representation for the ilmenite structure of Mn$_2$FeSbO$_6$}
\label{figilm}
\end{figure}

Single-crystals of mineral  Mn$_2$FeSbO$_6$ were obtained from the Swedish Museum of Natural History in Stockholm, Sweden. Several ceramic Mn$_2$FeSbO$_6$ samples were synthesized using a conventional solid state reaction (including a final sintering at temperatures between 1300 and 1350 $^{\rm o}$C), as well as using a thermobaric treatment. The thermobaric treatment was carried out under the pressure of 3 GPa for durations of 15-30 min at temperatures from 700 to 1000 $^{\rm o}$C (see Table~\ref{table-Psamples} for details). In each treatment, the pressure was first set to the rated value, and the temperature was increased. The sample was placed in a graphite heater lined with Pt foil to prevent interaction of the powder with the material of the heater.  On completion of the treatment, the sample was quenched to room temperature in less than 2 mins; after that the pressure was decreased to the normal. 

The phase composition of the prepared samples was studied by powder x-ray diffraction method on a D8 Bruker diffractometer using CuK$_{\alpha 1}$ radiation. The ICDD PDF4 database of standard powder patterns (ICDD, USA, Release 2009) was employed to identify possible impurity phases. The cation stoichiometry of the mineral and synthetic samples was checked using microprobe EDX analysis (average of 20 points). The magnetization of all samples was collected using a SQUID magnetometer MPMSXL from Quantum Design Inc. Several small ($\sim$ 10-500 $\mu$m) crystals of the mineral melanostibite were assembled together for magnetization measurements (total weight of 3.7 mg).\\

\begin{figure}[h]
\includegraphics{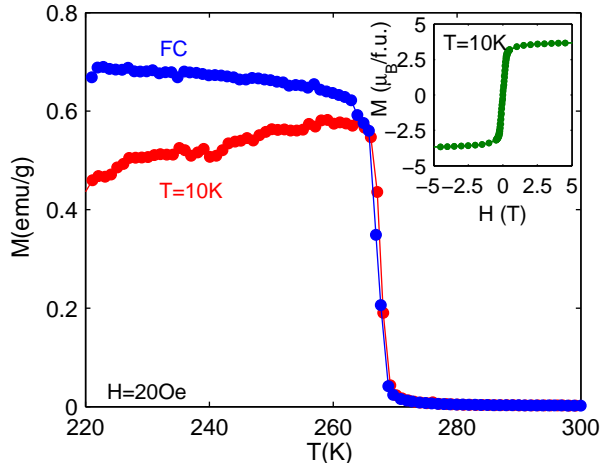}
\caption{(Color online) Main frame: Temperature dependence of the zero-field cooled (ZFC) and field cooled (FC) magnetization of the Mn$_2$FeSbO$_6$ mineral recorded in a small magnetic field ($H$ = 20 Oe). The inset shows the magnetic field dependence of the magnetization at $T$ = 10 K.}
\label{figmag1}
\end{figure}

Natural Mn$_2$FeSbO$_6$ crystallizes in the hexagonal ilmenite structure depicted in Fig.~\ref{figilm} (space group $R\bar{3}$, $a$=5.226 {\AA} and $c$= 14.325 {\AA}) with randomly distributed Fe/Sb cations.\cite{melano} In this structure, cations exist as Mn$^{2+}$ (3$d^5$), Fe$^{3+}$ (3$d^5$), and Sb$^{5+}$ (4$d^{10}$). Owning to their  3$d^5$ outer shell configurations, Mn$^{2+}$ and Fe$^{3+}$ are expected to carry 5 $\mu_B$ ($S$=5/2). As seen in Fig.~\ref{figmag1}, a sharp transition is observed near 270 K in the zero-field cooled / field cooled magnetization curves recorded  in a small magnetic field. At low temperatures the magnetization, which does not saturates, amounts in 5 Tesla to about  3.7 $\mu_B/f.u.$ . This suggests a ferrimagnetic arrangement in Mn$_2$FeSbO$_6$, with two moments in the formula unit parallel to each other, and the third one antiparallel.

\begin{figure}
\includegraphics{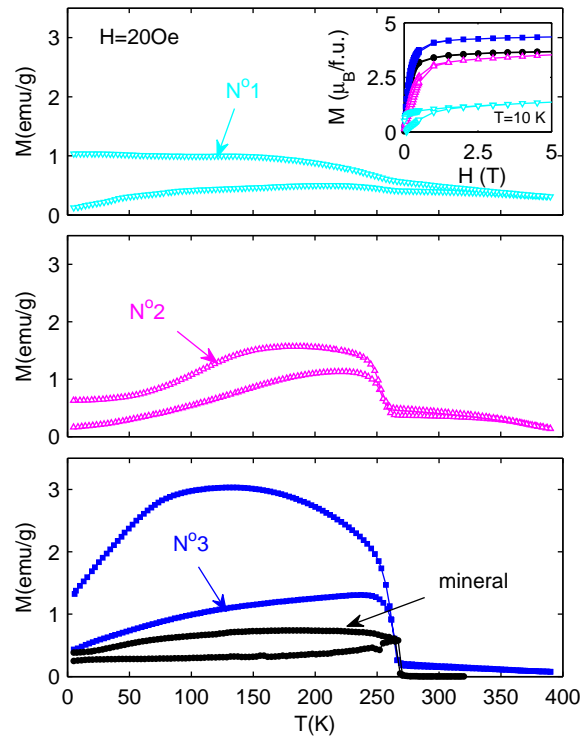}
\caption{(Color online) Temperature dependence of the zero-field cooled and field cooled magnetization recorded in a small magnetic field ($H$ = 20 Oe) for the synthetic Mn$_2$FeSbO$_6$ samples. The corresponding data for the mineral sample is added for comparison in the bottom panel. The inset shows the magnetic field dependence of the magnetization at $T$ = 10 K using the same symbols as in the panels.}
\label{figmag2}
\end{figure}

In order to elucidate the microscopic origin of such magnetic arrangement, we construct the low-energy electronic model for the $d$-states of ordered Mn$_2$FeSbO$_6$ alloy (model A) on the basis of first-principles electronic structure calculations, and further map it onto the Heisenberg model $\hat{H}=\sum_{i > j} J_{ij} {\bf S}_i \cdot {\bf S}_j$ with the spin $S=5/2$.\cite{review2008} As a test, we perform similar calculations for the hypothetical MnFeO$_3$ compound, where each Mn/Fe site was forced to accommodate five $d$-electrons (model B). Since superexchange interactions are defined separately for each pair of the Mn/Fe sites, these two models provide very similar description for the behavior of interatomic magnetic interactions. All interactions are antiferromagnetic and the magnetic system is frustrated.
Namely, the strongest interactions, which control the type of the magnetic arrangement, are the Mn-Fe interactions $J_{\rm Mn-Fe} \sim$ 2, 7, 5, and 4 meV between Mn and Fe atoms located in the first four coordination spheres. The Mn-Mn interactions are considerably weaker: $J_{\rm Mn-Mn} \sim$ 1 meV, which are restricted by the first coordination sphere. The Fe-Fe interactions in the first coordination sphere (derived in the model B) can be as large as $J_{\rm Fe-Fe} \sim$ 4 meV. However, the actual effect of these Fe-Fe interactions will be smaller due to the random distribution of the Fe/Sb atoms in Mn$_2$FeSbO$_6$. For example, due to this randomness, the Fe-Fe interactions will contribute to the total energy only with the weight 1/4. Thus, the magnetic configuration, corresponding to the minimum of the total energy, will be predetermined by the Mn-Fe interactions. In this configuration, Mn-Fe atoms are coupled antiferromagnetically along the $c$-axis, while Mn-Mn and Fe-Fe atoms across the inversion center are coupled ferromagnetically (i.e. akin to Fe$_2$O$_3$).\\

\begin{table*}[t]
\caption{Phase composition (ilmenite Mn$_2$FeSbO$_6$/pyrochlore Mn$_2$Sb$_2$O$_7$(MSO)/spinel MnFe$_2$O$_4$ (MFO) phases), and stoichiometry and structural parameters for Mn$_2$FeSbO$_6$ (MFSO) phase of the different samples; preparation conditions refer to temperature and time of thermobaric treatment at  3 GPa.\\}
\label{table-Psamples}

\begin{tabular}{c|c|c|c|c}
\colrule
&&&&\\
Sample&Preparation&Phase composition (\%)&Cation ratio for MFSO&Lattice parameters for\\
N$^{\rm o}$&conditions&MFSO/MSO/MFO&phase: Mn:Fe:Sb&MFSO phase: $a$, $c$({\AA})\\
&&&&\\
\colrule
&&&&\\
1&~700 $^{\rm o}$C for 20min.&82/13~/~5&1.93 : 1.13 : 0.94&5.243(1), 14.329(2)\\
2&~900 $^{\rm o}$C for 15min.&91/~5~/~4&1.95 : 1.08 : 0.97&5.238(1), 14.344(2)\\
3&1000 $^{\rm o}$C for 30min.&97/$<$1/~2&2.05 : 0.91 : 1.04 &5.237(1), 14.349(2)\\
Mineral& Nature & pure MFSO &2.03 : 0.96 : 0.99&5.226(1), 14.325(2)\\
&&&&\\
\colrule

\end{tabular}
\label{comp}
\end{table*}

The Mn$_2$FeSbO$_6$ samples synthesized using solid state reactions contain minor melanostibite phases, as well as secondary phases of Mn$_3$O$_4$\cite{mn3o4}, Mn$_2$Sb$_2$O$_7$\cite{mso227,mso227bis}, and MnFe$_2$O$_4$.\cite{mfo124} The magnetization of these samples (not shown) is dominated by the magnetism of the MnFe$_2$O$_4$ spinel, which is ferrimagnetic below $T_N$=570 K (about 3.27 $\mu_B/f.u.$ at 300 K).\cite{mfo124} On the other hand, as seen in the main frame of Fig.~\ref{figmag2}, the Mn$_2$FeSbO$_6$ samples fabricated under thermobaric treatment exhibit magnetic properties similar to those of the single-crystal mineral. Sample N$^{\rm o}$1 displays some magnetism up to the highest measured temperatures, albeit with a clear inflection in the magnetization near 270 K. X-ray diffraction and microstructure analysis revealed amounts of impurity phases of both Mn$_2$Sb$_2$O$_7$ pyrochlore and MnFe$_2$O$_4$ spinels (see Table~\ref{comp} for abundances). Hence the magnetism subsisting above 270 K is attributed to the presence of a MnFe$_2$O$_4$ phase. The Mn$_2$Sb$_2$O$_7$ pyrochlore is a frustrated magnet ordering below 15 K.\cite{mso227,mso227bis} As seen in Fig.~\ref{figmag2} and Table~\ref{comp}, the phase purity of the synthetic samples was greatly improved by increasing the temperature of the thermobaric treatment up to 1000$^{\rm o}$C. Magnetic hysteresis loops were also recorded at 300 K (not shown). For the mineral sample, only paramagnetic behavior was observed. However a spontaneous magnetic moment was observed in the synthetic sample, reflecting the presence of MnFe$_2$O$_4$ impurity (amounting to about 5 \% for sample N$^{\rm o}$1, and only 2 \% for N$^{\rm o}$3, in agreement with results from microstructure determination).

Sample N$^{\rm o}$3 contains minor amounts of secondary phases, and display a sharp magnetic transition near 270 K akin to that of the mineral sample. The magnitude of the magnetization at 5 Tesla also increases with the purity of the synthetic samples, reaching 4.35 $\mu_B/f.u.$ for the best sample (N$^{\rm o}$3). Interestingly, we have observed that if the pressure of the thermobaric treatment is increased to 6 GPa, Mn$_2$FeSbO$_6$ can be stabilized in a monoclinically distorted perovskite phase, with ordered Fe/Sb ions.\cite{bazuev} Preliminary magnetic studies suggest an antiferromagnetic transition below 60 K for that phase.\\

To summarize, we have investigated the magnetic properties of natural mineral and synthetic ceramic Mn$_2$FeSbO$_6$ with ilmenite structure.  By improving the preparation conditions, we could synthesize Mn$_2$FeSbO$_6$ with structure, phase purity and magnetism similar to those from the Nature. We believe that the sharp magnetic transition below room temperature suggests that functional magnetic materials based on Mn$_2$FeSbO$_6$ and its substitutions can be designed. Detailed studies of dielectric properties are underway.
 
\begin{acknowledgments}
We thank the Swedish Research Council (VR), the G\"oran Gustafsson Foundation, the Swedish Foundation for International Cooperation in Research and Higher Education (STINT), and the Russian Foundation for Basic Research for financial support. We are grateful to Dr. Henrik Skogby from the Swedish Museum of Natural History in Stockholm, Sweden, for providing melanostibite mineral. 
\end{acknowledgments}

\end{document}